\documentclass[aps,pre,reprint,superscriptaddress,showkeys]{revtex4-1}

\usepackage[utf8]{inputenc}
\usepackage{amsmath}
\usepackage{amssymb}
\usepackage{amsfonts}
\usepackage{graphicx}
\usepackage{psfrag}
\usepackage{siunitx}
\usepackage{color}

\newcommand{\fluc}{\mathrm{fluc}}
\newcommand{\diss}{\mathrm{diss}}
\newcommand{\avg}{\mathrm{avg}}
\newcommand{\eff}{\mathrm{eff}}

\renewcommand{\d}{\mathrm{d}}
\renewcommand{\Im}{\mathrm{Im}}

\begin{document}

\title{Low thermal fluctuations in a system heated out of equilibrium}

\author{Mickael Geitner}
\affiliation{Univ Lyon, Ens de Lyon, Univ Claude Bernard Lyon 1, CNRS, Laboratoire de Physique, F-69342 Lyon, France}

\author{Felipe Aguilar Sandoval}
\altaffiliation{Current address: Universidad de Chile, Facultad de Ciencias F\'isicas y Matem\'aticas, Departamento de F\'isica, Santiago, Chile}
\affiliation{Universidad de Santiago de Chile, Facultad de Ciencia, Departamento de F\'isica, Santiago, Chile}
\affiliation{Univ Lyon, Ens de Lyon, Univ Claude Bernard Lyon 1, CNRS, Laboratoire de Physique, F-69342 Lyon, France}

\author{Eric Bertin}
\affiliation{Universit\'e Grenoble Alpes and CNRS, LIPHY, F-38000 Grenoble, France}

\author{Ludovic Bellon}
\email{Corresponding author : ludovic.bellon@ens-lyon.fr}
\affiliation{Univ Lyon, Ens de Lyon, Univ Claude Bernard Lyon 1, CNRS, Laboratoire de Physique, F-69342 Lyon, France}

\date{\today}

\keywords{Fluctuations $|$ Out-of-equilibrium $|$ Fluctuation-Dissipation Theorem (FDT) $|$ Thermal noise $|$ Non Equilibrium Steady State (NESS)} 

\begin{abstract}
We study the mechanical fluctuations of a micrometer sized silicon cantilever subjected to a strong heat flow, thus having a highly non-uniform local temperature. In this non-equilibrium steady state, we show that fluctuations are equivalent to the thermal noise of a cantilever at equilibrium around room temperature, while its mean local temperature is several hundred of degrees higher. Changing the mechanical dissipation by adding a coating to the cantilever, we recover the expected rise of fluctuations with the mean temperature. Our work demonstrates that inhomogeneous dissipation mechanisms can decouple the amplitude of thermal fluctuations from the average temperature. This property could be useful to understand out-of-equilibrium fluctuating systems, or to engineer low noise instruments.
\end{abstract}

\maketitle

Brownian motion~\cite{Einstein-1905}, or more generally thermal fluctuations, are present in any system having a non-zero temperature, and affect all observables as a tiny random noise around their mean value. Meaningless for most macroscopic system, they become salient at the nano-scale at room temperature~\cite{Bustamante-2005}, in most elementary biological processes~\cite{Yanagida-2007}, or for high precision measurements, such as gravitational wave detection~\cite{Pitkin-2011}. In many cases, systems are out of equilibrium and the conclusions of the fluctuation-dissipation theorem cannot be applied~\cite{Marconi-2008,Cugliandolo-2011}: living systems~\cite{Kasas-2015,Dieterich-2015}, glasses~\cite{Cugliandolo-1997,Grigera-1999,Bellon-2001,Herisson-2002}, active matter~\cite{Loi-2011}, materials subject to a heat flux~\cite{Li-1994,Li-1998,Conti-2013}, chemical reactions~\cite{Santamaria-Holek-2011}, turbulent flows~\cite{Grenard-2008,Monchaux-2008} or shear fluids~\cite{Berthier-2002}, all present higher fluctuations than expected at equilibrium. In a few glassy systems~\cite{Berthier-2001,Mayer-2006,Gnan-2013,Berthier-2013}, fluctuations lower than those expected at equilibrium have however been predicted theoretically and observed numerically, but the non-stationary character of the dynamics is expected to play a key role in these observations.
In Non-Equilibrium Steady State (NESS), the expectation that fluctuations are enhanced with respect to equilibrium is further supported theoretically by the Harada-Sasa relation, which relates the excess of fluctuations to the energy dissipation rate in the framework of Langevin equations \cite{Harada-2005}.
Here we show experimentally and theoretically that in some cases, for a Non-Equilibrium Steady State (NESS) in a spatially extended system, thermal noise can present an apparent deficit with respect to equilibrium. We study the mechanical fluctuations of a micrometer sized silicon cantilever subjected to a strong heat flow, thus having a highly non-uniform temperature. We show that fluctuations are equivalent to the thermal noise of a cantilever at equilibrium around room temperature, while its mean temperature is several hundred of degrees higher. Our work, based on a generalization of the Fluctuation Dissipation Theorem (FDT), demonstrates that spatially inhomogeneous fields of temperature and dissipation mechanisms can lead to such an apparent deficit of thermal noise. This property could be useful to understand out of equilibrium fluctuating systems, or to engineer low noise instruments.

\begin{figure}[htbp]
\begin{center}
\includegraphics[width=75mm]{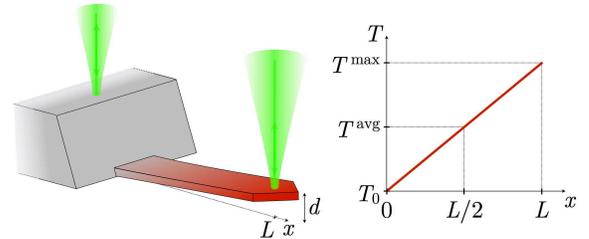}
\caption{Experiment principle: the deflection $d$ of a micro-cantilever (typically $\SI{500}{\micro m} \times \SI{100}{\micro m} \times \SI{1}{\micro m}$) is recorded through the interference of two laser beams, one reflected on the cantilever free end, the other on the chip holding the cantilever~\cite{Paolino-2013}. The sensing beam heats the cantilever and a steady temperature profile driven by the absorbed light power is reached, setting the system in a non-equilibrium steady-state. In a first approximation, the temperature $T$ grows linearly with $x$, the coordinate along the cantilever length~\cite{Aguilar-2015-JAP}.}
\label{fig:cantilever}
\end{center}
\end{figure}

\begin{figure}[htbp]
\begin{center}
\includegraphics{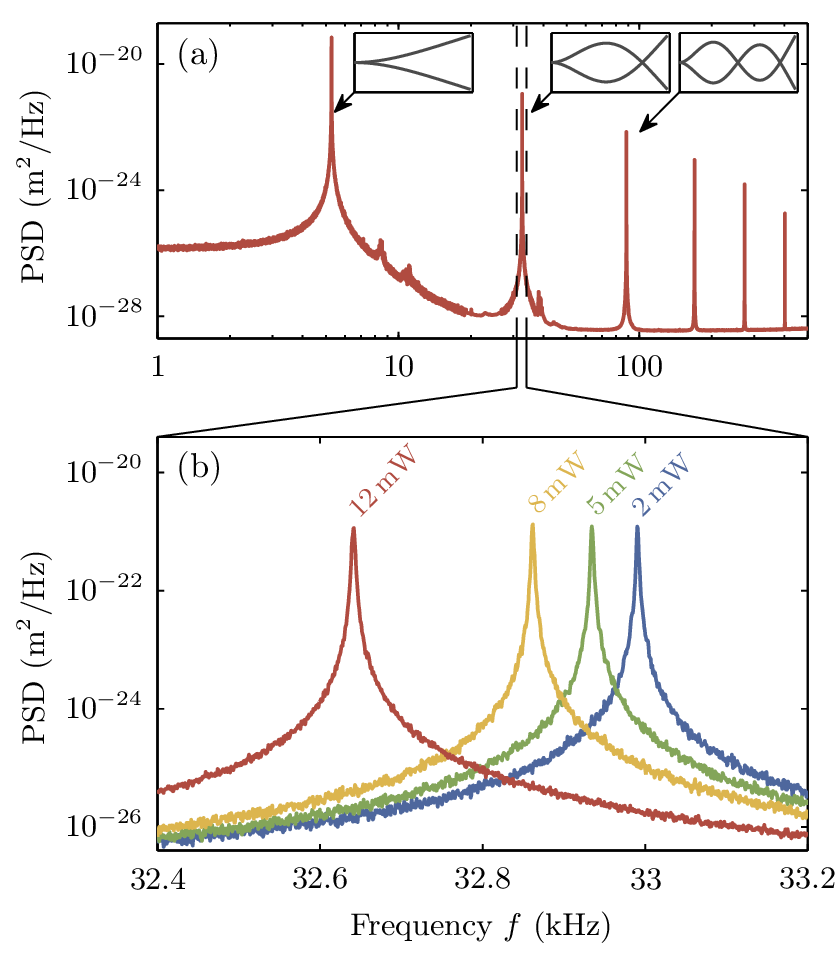}
\caption{Power spectrum density (PSD) of thermal noise induced deflection in vacuum as a function of frequency. (a) Each resonant mode is clearly identified by a sharp peak, associated to the normal mode shape pictured in the insets for the first 3 modes. These resonances can be modeled as uncoupled simple harmonic oscillators. The area under the PSD curve directly gives the mean square deflection of each of these degrees of freedom, and defines their effective temperatures. (b) When the light power grows, all the resonance frequencies of the cantilever decrease, as illustrated here with a zoom on mode 2. This frequency shift, due to the softening of the cantilever with temperature, is used to track the amplitude of the temperature profile, and thus the average temperature~\cite{Aguilar-2015-JAP}.}
\label{fig:PSD}
\end{center}
\end{figure}

The physical system studied is an atomic force microscopy cantilever heated by partial absorption of light from a laser focused at its free end, as illustrated in Fig.~\ref{fig:cantilever}. In vacuum, the main way out for the heat is by conduction~\cite{Note1} along the cantilever length towards the clamped base, which is thermalized at room temperature thanks to its macroscopic size. A huge thermal gradient can easily be reached: a $\SI{500}{K}$ difference along its $\SI{500}{\micro m}$ length with only $P=\SI{10}{mW}$ of light power~\cite{Aguilar-2015-JAP}. We measure the deflection with a differential interferometer~\cite{Paolino-2013}, whose sensing beam also acts as the heating source. More details about the experimental setup are given in appendix~\ref{appendixA}. A typical power spectrum density (PSD) of the deflection is plotted in Fig.~\ref{fig:PSD}(a): each sharp resonance is easily identified as a flexural mode of oscillation driven by thermal noise only. In vacuum, the quality factor of these resonances is several thousands, each one can thus be seen as an independent degree of freedom modeled by an harmonic oscillator~\cite{Butt-1995}. Since silicon Young's modulus decreases with temperature, the cantilever gets slightly softer when heated, thus the resonance frequencies decrease. Tracking those frequency shifts with light power, we estimate the cantilever temperature profile, and thus its average temperature $T^{\avg}$ (more details are available in appendix~\ref{appendixB} and in ref.~\cite{Aguilar-2015-JAP}).

Although from a thermal point of view, the cantilever can be described within the framework of local equilibrium with a temperature profile $T(x)$, the fluctuations of the deflection of the cantilever are a collective phenomenon, that cannot be described only by the local temperature at the endpoint of the cantilever; these fluctuations rather need to be described using the normal modes of the cantilever. The present experimental system then appears as an ideal toy model to test the behavior of the thermal fluctuations in a Non-Equilibrium Steady State (NESS): the system is large enough to have a strongly non-uniform temperature profile, and small enough to have measurable mechanical fluctuations, its degrees of freedom (i.e., the normal modes) are easily identified and uncoupled, and a one-dimensional description (the Euler-Bernoulli model for a mechanical clamped-free beam, see below) captures the main physical phenomena . 

To quantify the amplitude of the thermal fluctuations of each mode $n$, we define an effective temperature $T_{n}^{\eff}$ by an extended equipartition relation:
\begin{equation} \label{eq:Teff}
\frac{1}{2}k_{n} \langle d_{n}^{2} \rangle = \frac{1}{2}k_{B} T_{n}^{\eff},
\end{equation}
where $k_{n}$ is the stiffness of the mode, $k_{B}$ the Boltzmann constant, and $\langle d_{n}^{2} \rangle$ the mean square deflection of mode $n$, measured by integrating the power spectrum density of thermal noise on a small frequency range around the resonance (see appendix~\ref{appendixC} for details). We plot in Fig.~\ref{fig:Tn} the effective temperatures of the first 3 flexural modes measured on two different cantilevers as a function of the impinging light power $P$, and compare it to their average temperature $T^{\avg}$ (measured by the resonance frequency shift~\cite{Aguilar-2015-JAP}). The behavior of the two cantilevers is strikingly different. For the raw silicon one, $T_{n}^{\eff}$ is roughly mode independent and almost constant, close to the room temperature $T_{0}=\SI{295}{K}$. The system thus presents a \emph{deficit of fluctuations} with respect to what would be expected for a system in equilibrium around its average temperature.
Even more strikingly, the resulting effective temperature $T_{n}^{\eff}$ is also much lower than any simple weighted average of the temperature profile $T(x)$ that could have been anticipated, like an average of $T(x)$ weighted by the amplitude (or the squared amplitude) of the mode for instance.
For the tantala coated cantilever, $T_{n}^{\eff}$ is mode dependent, but it increases with the light power $P$ and stands reasonably close to the average temperature for $n>1$. However, the first mode also seems to present a deficit of fluctuations with respect to an equivalent equilibrium situation. 

\begin{figure}[htbp]
\begin{center}
\includegraphics{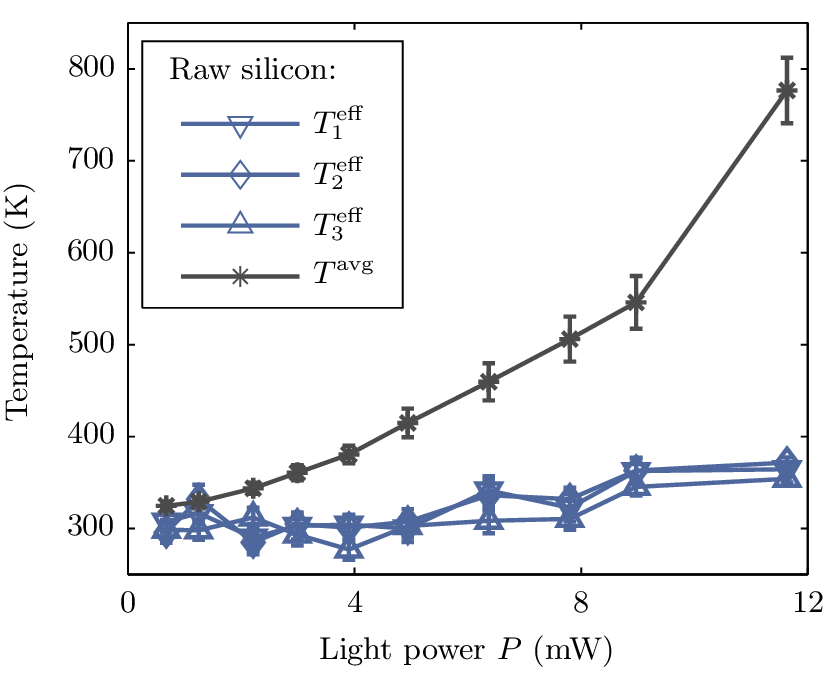}
\includegraphics{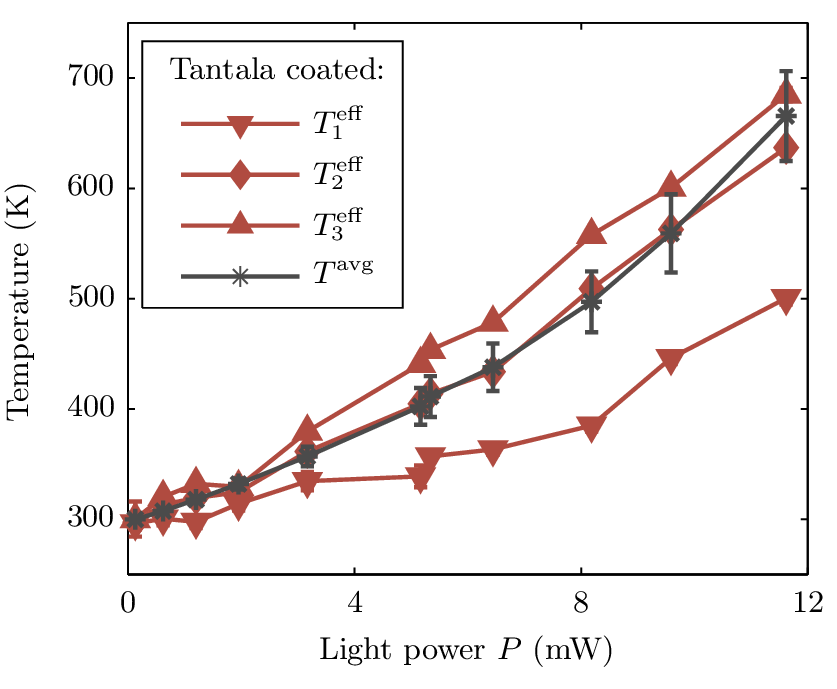}
\caption{Effective temperatures of the first 3 flexural modes ($T_{n}^{\eff}$) and average temperature (stars, $T^{\avg}$) measured on two different cantilevers as a function of the impinging light power $P$. (top) Raw silicon cantilever:  $T_{n}^{\eff}$ is mode independent and almost constant (equal to room temperature), when $T^{\avg}$ raises by $\SI{400}{K}$. (bottom) tantala coated silicon cantilever: $T_{n}^{\eff}$ is mode dependent, and increases with $P$ similarly to $T^{\avg}$ for $n>1$. Error bars (one standard deviation) correspond to the statistical uncertainty for $T_{n}^{\eff}$ (at most $\SI{15}{K}$), and to the systematic error for $T^{\avg}$ (discrepancy between the evaluation on the first 3 modes). For both cantilevers, measurable higher order modes present the same behavior, but their amplitude being smaller uncertainty increases and we do not present them for sake of clarity.}
\label{fig:Tn}
\end{center}
\end{figure}

To interpret those unexpected observations, let us first note that $T_{n}^{\eff}$ is defined for an spatially extended system in vacuum. Thus, the extended equipartition theorem (eq. \ref{eq:Teff}) for a global deflection degree of freedom is qualitatively different from the usual setting where an harmonic degree of freedom is coupled to a thermal bath that would dictate its temperature. Let us investigate a one-dimensional model of the cantilever, namely the
Euler-Bernoulli model for a mechanical clamped-free beam.
In this framework, the cantilever is described by the one-dimensional deflection field $d(x,t)$ at time $t$, with $x$ the dimensionless spatial coordinate along the beam ($0<x<1$), which is clamped at $x=0$.
It is convenient to work with the time-Fourier transform $d(x,\omega)$ of the deflection field, which is described by~\cite{Aguilar-2015-JAP}
\begin{equation} \label{eq:GEB}
-m \omega^{2} d(x,\omega) + \mathcal{L} d(x,\omega) = F(x,\omega)
\end{equation}
with $F(x,\omega)$ the external force acting on the cantilever.
The operator $\mathcal{L}=\mathcal{L}^{r}+i\mathcal{L}^{i}$ can be written in vacuum as 
\begin{equation} \label{eq:L}
\mathcal{L}=\frac{\partial^{2}}{\partial x^{2}}\left([k^{r}(x,\omega) + ik^{i}(x,\omega)]\frac{\partial^{2}}{\partial x^{2}}\right)\end{equation}
with $k^{r}(x,\omega)$ and $k^{i}(x,\omega)$ respectively the local stiffness and viscoelasticity (internal dissipation). 
At equilibrium at temperature $T$, the cantilever satisfies the fluctuation-dissipation relation
\begin{equation} \label{eq:etaxetaxp}
S_{F}(x,x',\omega) = \frac{2}{\pi}\frac{k_{B}T}{\omega} \frac{\partial^2}{\partial x^2} \left[ k^{i}(x,\omega)\frac{\partial^2}{\partial x^2} \delta(x-x') \right].
\end{equation}
where $S_{F}(x,x',\omega)$ is the PSD of the force $F$, defined from the relation
\begin{equation}
\langle F(x,\omega)F(x',\omega') \rangle = S_{F}(x,x',\omega) \, \delta(\omega+\omega')
\end{equation}
Assuming that fluctuations at point $x$ are described by the temperature field $T(x)$, and taking into account the local conservation of momentum that imposes that the PSD has to be written as a second derivative in space \cite{Dean-1996}, a minimal generalization of Eq.~(\ref{eq:etaxetaxp}) to the nonequilibrium situation reads
\begin{equation} \label{eq:FDR-outofeq}
S_{F}(x,x',\omega) = \frac{2}{\pi}\frac{k_{B}}{\omega}
\frac{\partial^2}{\partial x^2} \left[ T(x) k^{i}(x,\omega) \frac{\partial^2}{\partial x^2} \delta(x-x') \right]
\end{equation}
Expanding the deflection field $d(x,\omega)$ over the normal modes, and integrating Eq.~(\ref{eq:FDR-outofeq}) over frequency yields the mean square deflection 
$\langle d_{n}^{2} \rangle$ of mode $n$.
The corresponding effective temperature $T_{n}^{\eff}$, as defined in
Eq.~\ref{eq:Teff}, is then obtained as an average of the temperature profile $T(x)$, weighted by the normalized local rate $w_{n}^{\diss}(x)$ of energy dissipation associated with mode $n$:
\begin{equation} \label{eq:Tn}
T_{n}^{\eff} = \int_{0}^{1} T(x) w_{n}^{\diss}(x) \d x.
\end{equation}
(see appendix~\ref{appendixD} for details on the derivation). The normalized dissipation rate $w_{n}^{\diss}(x)$ is defined as
\begin{equation} \label{eq:wdiss:norm}
w_n^{\rm diss}(x) =
\frac{\phi_n''(x)^2 k^{i}(x,\omega_n)}{\int_0^1 \d x' \, \phi_n''(x')^2 k^{i}(x',\omega_n)}
\end{equation}
where $\phi_n''(x')$ is the local curvature of mode $n$. In other words, the amplitude of fluctuations, quantified by the effective temperature $T_{n}^{\eff}$, is on average proportional to the temperature at the position where mechanical energy is dissipated.

Raw silicon cantilevers for example, that are manufactured by chemical etching from a single cristal silicon wafer, are virtually free of defects, and very little internal damping is expected along their length. Indeed, if the sole dissipation process was the thermoelastic damping (coupling between thermal phonons and mechanical vibrations), the quality factor of the resonances would be of the order of $\num{E6}-\num{E7}$~\cite{Lifshitz-2000}, much higher than what we measure around $\num{E5}$. In vacuum, viscous damping is not relevant either~\cite{Li-2012}, hence most mechanical dissipation should occur around the clamped base, which is at room temperature $T_{0}$: $w_{n}^{\diss}(x) \sim \delta(x)$. And indeed, measured fluctuations are constant with $T_{n}^{\eff}\sim T_{0}$, whatever the average temperature of the cantilever.

To further assess our approach, we now focus our attention to the tantala coated cantilever. The amorphous coating introduces a viscoelastic damping, almost frequency independent, along the cantilever~\cite{Paolino-2009-Nanotech,Li-2012}. The quality factor of the resonances indeed drops from $\num{E5}$ for the raw silicon sample to $\num{2E3}$ for this coated sample~\cite{Li-2014}. If this damping is supposed to be uniform along the cantilever, then the normalized local rate of energy dissipation is simply proportional to the square of the normal mode curvature: $w_{n}^{\diss}(x)\propto \phi''_{n}(x)^{2}$ (see appendix~\ref{appendixD} material for details). Let us assume a linear temperature profile $\Delta T(x)=2 x \Delta T^{\avg}$~\cite{Note2}, as expected from the Fourier law for a constant thermal conductivity~\cite{Aguilar-2015-JAP}. It is then easy to compute a simple coefficient $\kappa_{n}$ linking the effective temperature rise of mode $n$ to $\Delta T^{\avg}$:

\begin{align}
\Delta T_{n}^{\eff} &= \kappa_{n} \Delta T^{\avg},\\
\kappa_{n}&=\frac{1}{\int_{0}^{1} \phi''_{n}(x)^{2} \d x}\int_{0}^{1} 2 x \phi''_{n}(x)^{2} \d x. \label{eq:kappa_n}
\end{align}
Numerical application leads to $\kappa_{1}=0.39$, $\kappa_{2}=0.81$, $\kappa_{3}=0.94$ and  $\kappa_{n\rightarrow \infty} \rightarrow 1$. The first mode dissipates most of its mechanical energy at the origin, where the highest curvatures are located, and thus present lower fluctuations than a system in equilibrium at the same average temperature. For higher order modes, energy dissipation tends to be distributed all cantilever long, thus $T_{n}^{\eff}$ tends to the average temperature. Those two features are indeed observed for the tantala coated cantilever in Fig.~\ref{fig:Tn}.

\begin{figure}[htbp]
\begin{center}
\includegraphics{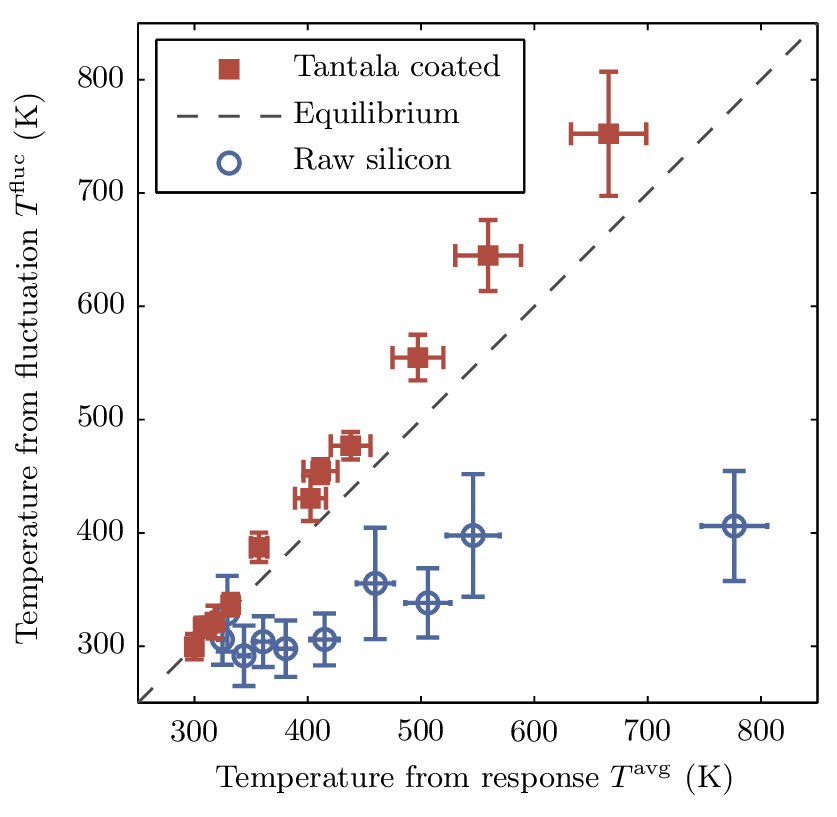}
\caption{Average temperatures : $T^{\fluc}$, deduced from fluctuations, versus $T^{\avg}$, deduced from the frequency shift of the resonances.  $T^{\fluc}$ quantifies the thermal fluctuations amplitude: it is the average on the first 3 modes of the effective temperatures $T_{n}^{\eff}$ rescaled for the mode shape through coefficient $\kappa_{n}$, thus assuming a uniform mechanical damping along the cantilever. The raw silicon cantilever presents a strong deficit of fluctuations with respect to equilibrium, while the tantala coated one presents a small excess of fluctuations. Error bars (one standard deviation) accounts for the statistical uncertainty and systematic error (discrepancy between the evaluation on the first 3 modes).}
\label{fig:Tavg}
\end{center}
\end{figure}

In this framework, we compare in Fig.~\ref{fig:Tavg} the average temperature estimated from the fluctuations of the first 3 modes,
\begin{equation}
\Delta T^{\fluc}=\frac{1}{3}\sum_{n=1}^{3} \frac{\Delta T_{n}^{\eff}}{\kappa_{n}},
\end{equation}
to the one measured from the response of the cantilever $T^{\avg}$, deduced from the frequency shift of the  resonances~\cite{Note3}. The error bars in this figure take into account the statistical uncertainty and the dispersion of the estimations between the first 3 modes. For the tantala coated cantilever, the fluctuations are slightly in excess with respect to equilibrium. This maximum $\SI{20}{\%}$ excess can be due to the NESS, but could also come from a departure of the uniform dissipation hypothesis: hotter parts of the cantilever have higher mechanical losses, thus an increased weight in the estimation of $T_{n}^{\eff}$ in eq.~(\ref{eq:Tn}). For the raw silicon cantilever, the strong deficit of fluctuations is salient again. Moreover, the large error bars show a discrepancy between the temperatures estimated from the different modes. The hypothesis of uniform dissipation thus does not apply to this cantilever, whose behavior is much better understood with a localized dissipation around the clamped base.

Our observations contrasts singularly with the experiment reported in \cite{Conti-2013}, in which thermal fluctuations of a mass attached to a clamped rod subjected to a heat flux have an amplitude not only larger that those expected from the average temperature, but even larger than those expected from the maximum temperature. The experiment and framework proposed here clearly demonstrate that some non-equilibrium situations can on the contrary lead to unexpectedly low thermal fluctuations, with respect to a system in equilibrium having the same average temperature: the amplitude of the fluctuations can be prescribed by only the lowest temperature of a spatially extended system, when mechanical dissipation is localized in the low temperature area of the system. Large fluctuations corresponding to the highest temperature could just as well be observed: in general, one has to consider how the spatial distribution of dissipation weights the temperature field to predict the overall fluctuations of the system from the proposed extension of the fluctuation-dissipation relation.

Our findings could have many practical applications. For example, next generation gravitational wave detectors~\cite{Pitkin-2011} aim at lowering the detection of thermal noise by cryogenic operation: our results suggest that the temperature where mechanical dissipation occur should be the focus of attention. Owing to the generality of the underlying statistical mechanism, those findings should be independent of the actual system: mechanical devices, electronic circuits or biological systems could present equivalent behaviors in some non-equilibrium steady state situations.

\begin{acknowledgments}
We thank F. Vittoz and F. Ropars for technical support, A. Petrosyan, S. Ciliberto for stimulating discussions, and V. Dolique, R. Pedurand and G. Gagnoli of the Laboratoire de Mat\'eriaux avanc\'es in Lyon for the tantala coating of one of the cantilevers. This work has been supported by the ERC project \emph{OutEFLUCOP} and the ANR project \emph{HiResAFM} (ANR-11-JS04-012-01) of the Agence Nationale de la Recherche in France.
\end{acknowledgments}

\appendix

\section{Experimental setup}\label{appendixA}

The experiments are performed in vacuum (pressure below $\SI{e-2}{mbar}$) on two different cantilevers: a raw silicon cantilever (Nanoworld Arrow TL8), $\SI{500}{\micro m}$ long, $\SI{100}{\micro m}$ wide and $\SI{1}{\micro m}$ thick, and a tantala coated silicon cantilever (BudgetSensors AIO-TL), $\SI{500}{\micro m}$ long, $\SI{30}{\micro m}$ wide, $\SI{2.7}{\micro m}$ thick, with a $\SI{300}{nm}$ $\mathrm{Ta_{2}0_{5}}$ coating on each side of the cantilever~\cite{Li-2014}. This tantala coating was performed by the LMA (Laboratoire des Matériaux Avanc\'es, Lyon, France) using ion beam sputtering, and was annealed to relax the internal stresses and to minimize the cantilever static curvature.

The thermal noise induced deflection is measured with a home made quadrature phase differential interferometer~\cite{Paolino-2013}: the optical path difference between a sensing beam focused on the free end of the cantilever and a reference beam focused on the chip holding the cantilever gives a direct and calibrated measurement of the deflection. We acquire the signals from the photodiodes of the interferometer with high resolution DAQ cards from National Instruments (NI-PXI 5922). From the recorded signals (typically $\SI{100}{s}$ sampled at $\SI{1}{MHz}$), we compute the deflection $d$, then its power spectrum density $S_{d}$ to estimate the thermal noise spectrum. The measurement is only shot noise limited as illustrated in Fig.~\ref{fig:PSD}(a): the thermal noise of each resonance clearly stand above the shot noise limit (background white noise around $\SI{e-28}{m^{2}/Hz}$). We limit our analysis to the first 3 modes on the two cantilevers, where the signal to noise ratio is large enough for all laser powers used. For both cantilevers, measurable higher order modes present the same behavior as mode 3, but their amplitude being smaller uncertainty increases and we do not present them for sake of clarity.

The laser of the interferometer is also used as the heat source to drive the cantilever in a NESS. We use a single mode, continuous wave diode pumped solid state laser (Spectra-Physics Excelsior), with up to $\SI{150}{mW}$ at $\SI{532}{nm}$. The light power $P$ is controlled using neutral density filters at the output of the laser, ensuring optimal stability (better than $\SI{2}{\%}$). Before and after any thermal noise acquisition, we translate the laser sensing beam out of the cantilever, then measure $P$ with a calibrated photodiode below the cantilever in the vacuum chamber.

\section{Average temperature measurement from frequency shift}\label{appendixB}

Each resonant mode $n$ of the cantilever can be modeled by a simple harmonic oscillator: its resonance angular frequency $\omega_{n}$ is linked to the normal mode stiffness $k_{n}$ and effective mass $m$ (one fourth of the cantilever mass) through the usual relation $k_{n}=m \omega_{n}^{2}$. $k_{n}$ is proportional to the Young's modulus of silicon (or a combination of that of silicon and tantala), which decreases with temperature. The resonant modes are thus shifted to a lower frequency when the temperature rises~\cite{Aguilar-2015-JAP}, as illustrated in Fig.~\ref{fig:PSD}(b). 

If the cantilever temperature rise $\Delta T$ is uniform, then the frequency shift is simply given by
\begin{equation} \label{eq:DfSHO}
\frac{\Delta \omega_{n}}{\omega_{n}} = \frac{1}{2} \frac{\Delta k_{n}}{k_{n}} = \frac{1}{2} \frac{\Delta E}{E} = \frac{1}{2} \alpha_{E} \Delta T,
\end{equation}
where $\alpha_{E}=(1/E) \partial_{T} E$ is the temperature coefficient of the Young's modulus of silicon. For our two samples, we directly measure $\alpha_{E}$: in vacuum, we change the global temperature of the cantilever holder from $T_{0}$ to $\SI{120}{\degree C}$ and use a small laser power to avoid additional heating; we then track the relative frequency shift $\Delta \omega_{n}/\omega_{n}$ as a function of the imposed $\Delta T$. We check that the result is mode independent and get the following results : $\alpha_{E}=(-64\pm 2)\times\SI{e-6}{K^{-1}}$ for raw silicon (in agreement with tabulated values~\cite{Bourgeois-1997}), and $\alpha_{E}=(-113\pm 4)\times\SI{e-6}{K^{-1}}$ for the tantala coated cantilever.

For a localized heating by laser absorption, one should consider the temperature profile in the cantilever. Under the linear approximation hypothesis $\Delta T(x)=\Delta T^{\avg} \, 2 x / L$, the frequency shift is mode dependent and reads~\cite{Aguilar-2015-JAP}:
\begin{equation}
\frac{\Delta \omega_{n}}{\omega_{n}} = \frac{1}{2} \kappa_{n} \alpha_{E} \Delta T^{\avg},
\end{equation}
where $\kappa_{n}$ is defined by eq.~(\ref{eq:kappa_n}). Indeed, in the global frequency shift, the local contribution of the cantilever softening is weighted by $\alpha_{E} \Delta T (x)$ times the square of local curvature $\phi''_{n}(x)^{2}$ of the considered mode. $\alpha_{E}$ being calibrated and $\kappa_{n}$ numerically computed, we therefore get for each mode (here the first 3 ones) an estimation of the cantilever average temperature, and average them together to estimate $\Delta T^{\avg}$: 
\begin{equation}
\Delta T^{\avg} =  \frac{1}{3} \sum_{n=1}^{3} \frac{2}{\alpha_{E}\kappa_{n}} \frac{\Delta \omega_{n}}{\omega_{n}}.
\end{equation}
The error bars plotted for $T^{\avg} $ on Figs.~\ref{fig:Tn} and \ref{fig:Tavg} correspond to the standard deviation between the estimations on the first 3 modes, they characterize the systematic error due to the approximation of the linear temperature profile. Due to the sharp nature of the resonant peaks, statistical error is negligible in the frequency shift measurement. A last source of error could come from the dependence on temperature of the Young's modulus at high temperature, but no strong departure from the linear coefficient $\alpha_{E}$ is expected~\cite{Bourgeois-1997}.

\section{Effective temperature from thermal noise measurement}\label{appendixC}

The effective temperature of each mode $T_{n}^{\eff}$ is defined by eq.~(\ref{eq:Teff}). Its measurement thus relies on that of the mode stiffness $k_{n}$ and of the mean square deflection $\langle d_{n}^{2} \rangle$. The latter is estimated by integrating the power spectrum density $S_{d}$ of thermal noise on a small frequency range around the resonance, taking care of subtracting the background noise contribution. Since the quality factor of each resonance is huge, the actual integration range has very little influence (below $\SI{1}{\%}$) on the result.

The stiffness $k_{n}$ is estimated by extrapolating the thermal noise at zero laser power $P$. First we perform a linear fit of $\langle d_{n}^{2} \rangle$ as a function of $P$ for $P<\SI{4}{mW}$, and take the ordinate at the origin as an estimation of the mean square deflection in absence of heating. This point corresponds to a system in equilibrium at temperature $T_{0}$, thus the Fluctuation-Dissipation Theorem (FDT) applies and the value of $k_{n}$ can be computed. The minute changes to $k_{n}$ for higher laser powers are then estimated from the frequency shift measurement, using eq.~(\ref{eq:DfSHO}).

The estimation of $T_{n}^{\eff}$ relies on measuring the mean square value of a random observable, and is thus subject to a statistical uncertainty. To estimate this uncertainty, we measure $T_{n}^{\eff}$ on typically $N=30$ time windows of $\SI{3}{s}$ each, and compute the standard deviation $\sigma_{n}$ of these $N$ measurements. The statistical error bars plotted in Fig.~\ref{fig:Tn} are then estimated by $\sigma_{n}/\sqrt{N}$, they are barely visible (below $\SI{20}{K}$ at worse).

In Fig.~\ref{fig:Tavg}, the error bars also take into account the dispersion between the first 3 modes (standard deviation among the 3 estimations), they characterize the systematic error due to the approximation of the linear temperature profile and uniform mechanical dissipation. Note however that if only the latter assumption is made, any departure from the former hypothesis is equivalent on both axis: indeed, in both cases the temperature profile $\Delta T(x)$ is weighted by the same quantity (the square of local curvature $\phi''_{n}(x)^{2}$ of the considered mode). 

\section{Extended FTD}\label{appendixD}

The rectangular cantilever is described in an Euler Bernoulli framework: its length $L$ is supposed to be much larger than its width, which itself is much larger than its thickness. We will limit ourselves in this study to the flexural modes of the cantilever: the displacement field is supposed to be only perpendicular to its main plane and uniform across its width. These deformations can thus be described by the deflection $d(x,t)$, $x$ being the normalized spatial coordinate along the beam, and $t$ the time. In Fourier space, the equation describing the evolution of the deflection $d(x,\omega)$ reads~\cite{Aguilar-2015-JAP}
\begin{equation}
-m \omega^{2} d(x,\omega) + \mathcal{L} d(x,\omega) = F(x,\omega)
\end{equation}
with $F(x,\omega)$ the external force acting on the cantilever, and where the operator $\mathcal{L}=\mathcal{L}^{r}+i\mathcal{L}^{i}$ can be written in vacuum as 
\begin{equation} \label{eq:LAppendix}
\mathcal{L}=\frac{\partial^{2}}{\partial x^{2}}\left([k^{r}(x,\omega) + ik^{i}(x,\omega)]\frac{\partial^{2}}{\partial x^{2}}\right)\end{equation}
with $k^{r}(x,\omega)$ and $k^{i}(x,\omega)$ respectively the local stiffness and viscoelasticity (internal dissipation). Note that in general the operator $\mathcal{L}$ can depend on $\omega$, though we will always suppose this variation
to be slow: it can be considered constant around each resonance of the cantilever.

Following Refs.~\citenum{Callen-1951} and \citenum{Levin-1998}, $d(x,t)$ and $F(x,t)$ are coupled variables by the Hamiltonian of the system ($x$ being a parameter), thus we may apply the FDT using these variables when the system is in equilibrium. We will use the FDT in Fourier space for the two observables $F(x,\omega)$ and $F(x',\omega)$ at two different position $x$ and $x'$. The displacement $d(x',\omega)$ plays the role of the ``field'' conjugated to $F(x',\omega)$, so that one can write the following fluctuation-dissipation relation,
\begin{equation} \label{eq:FDT}
S_{F}(x,x',\omega)=\frac{2}{\pi}\frac{k_{B}T}{\omega} \Im\big( R(x,x',\omega) \big)
\end{equation}
where $S_{F}(x,x',\omega)$ is the PSD of the force F and  $R(x,x',\omega)$ the response function of $F(x,\omega)$ to a field $d(x',\omega)$ applied at position $x'$. $S_{F}$ and $R$ are mathematically defined by

\begin{align}
\langle F(x,\omega)F(x',\omega') \rangle &= S_{F}(x,x',\omega) \, \delta(\omega+\omega') \label{eq:def:Sxx} \\
R(x,x',\omega) &= [-m\omega^2+\mathcal{L}] \, \delta(x-x') \label{eq:def-Y}
\end{align}
Taking the imaginary part of Eq.~(\ref{eq:def-Y}) yields
\begin{equation}
\Im\big( R(x,x',\omega) \big) = \mathcal{L}^{i} \delta(x-x')
\end{equation}
Using the explicit form of $\mathcal{L}^{i}$ given in eq.~(\ref{eq:LAppendix}), the fluctuation-dissipation relation (\ref{eq:FDT}) can be rewritten as
\begin{equation} \label{eq:etaxetaxpAppendix}
S_{F}(x,x',\omega) = \frac{2}{\pi}\frac{k_{B}T}{\omega} \frac{\partial^2}{\partial x^2} \left[ k^{i}(x,\omega)\frac{\partial^2}{\partial x^2} \delta(x-x') \right].
\end{equation}

If the cantilever is subject to a heat flux, its temperature is not uniform anymore. We wish to generalize eq.~(\ref{eq:etaxetaxpAppendix}) by adding the dependence of $T$ on space. We have to take into account the fact that due to momentum conservation, the noise $F(x,\omega)$ is a conserved noise, so that its correlation has to be expressed as a second derivative~\cite{Dean-1996}. The generalization of eq.~(\ref{eq:etaxetaxpAppendix}) to a nonuniform temperature profile then reads
\begin{equation}
S_{F}(x,x',\omega) = \frac{2}{\pi}\frac{k_{B}}{\omega}
\frac{\partial^2}{\partial x^2} \left[ T(x) k^{i}(x,\omega) \frac{\partial^2}{\partial x^2} \delta(x-x') \right]
\end{equation}

We now focus our attention on the resonant modes of the cantilever, whose spatial shape is given by the eigenvectors $\phi_{n}(x)$ of $\mathcal{L}^{r}$ associated to the eigenvalue $k_n^{r}$:
\begin{equation}
\mathcal{L}^{r} \phi_{n}(x) = k_{n}^{r} \phi_{n}(x).
\end{equation}
Those normal modes are described by the deflection amplitude $d_{n}(\omega)$ and conjugated force $F_{n}(\omega)$. The computation of the power spectral density of $F_{n}$ then yields

\begin{align} \nonumber
S_{F_{n}} (\omega) 
&=  \int_{0}^{1} \d x \int_{0}^{1} \d x' \, \phi_n(x) \phi_n(x')
S_{F}(x,x',\omega)\\
&= \frac{2}{\pi} \frac{k_{B}}{\omega}\int_{0}^{1} \d x\, \phi''_{n}(x)^2 T(x) k^{i}(x,\omega)
\end{align}

The mean square deflection of mode $n$ is given by
\begin{align}
\langle \d_{n}^{2} \rangle & = \int_{0}^{\infty} \d\omega \left| \frac{d_{n}(\omega)}{F_{n}(\omega)}\right|^{2} S_{F_{n}}(\omega)\\
& = \frac{2}{\pi} k_{B} \int_{0}^{\infty} \frac{\d\omega}{\omega} \frac{\int_0^1 \d x \, \phi_n''(x)^2 T(x) k^{i}(x,\omega)}{(k_{n}^{r}(\omega)-m\omega^{2})^{2}+k_{n}^{i}(\omega)^{2}}  
\end{align}
where the internal damping $k_{n}^{i}(\omega)$ of mode $n$ is given by
\begin{equation} \label{eq:gamman:eq}
k_{n}^{i}(\omega) = \int_0^1 \d x \, \phi_n''(x)^2 k^{i}(x,\omega).
\end{equation}
The spectrum being sharply peaked around $\omega_{n}$, the slow dependence of $k_{n}^{r}$ and $k_{n}^{i}$ on $\omega$ can be neglected, the mean square deflection then reads
\begin{equation}
\langle d_{n}^{2} \rangle \approx 
\frac{k_B}{k^{r}_{n}(\omega_n)}\frac{1}{k^{i}_{n}(\omega_n)} \int_0^1 \d x \,\phi_n''(x)^2 T(x)  k^{i}(x,\omega_n).
\end{equation}
In equilibrium, ones recovers the classic equipartition result:
\begin{equation}
\langle d_{n}^{2} \rangle = \frac{k_B T}{k_n^{r}}
\end{equation}
By analogy, we defined the mode effective temperature with eq.~(\ref{eq:Teff}), which leads to the following expression for $T_n^{\eff}$:
\begin{equation}
T_n^{\eff} = \int_0^1 \d x \,T(x) w_n^{\rm diss}(x)
\end{equation}
where we have defined the local, normalized, dissipation rate
\begin{equation} \label{eq:wdiss:normAppendix}
w_n^{\rm diss}(x) =
\frac{\phi_n''(x)^2 k^{i}(x,\omega_n)}{\int_0^1 \d x' \, \phi_n''(x')^2 k^{i}(x',\omega_n)}
\end{equation}
in agreement with Eq.~(\ref{eq:Tn}).
The mode-dependent temperature computed from this formula is thus a spatial average of the temperature profile, weighted by the local dissipation of the considered mode.

To show that $w_n^{\rm diss}(x)$ can indeed be interpreted as a local dissipation, let us assume that $k^{i}(x,\omega)=\gamma(x)\omega$ and that $k^{r}$ is frequency independent (which are reasonable approximations around the sharp resonances of the cantilever).
It is then easier to work in the time domain instead of Fourier space.
The Euler-Bernouilli equation~of motion for the cantilever is given by:
\begin{equation} \label{eqmotion1}
m \frac{\partial^{2} d}{\partial t^{2}} + \frac{\partial^{2}}{\partial x^{2}}
\left[ \left( k^{r}(x)+\gamma(x) \frac{\partial}{\partial t}\right) 
\frac{\partial^{2} d}{\partial x^{2}}\right] = F(x,t)
\end{equation}
To determine power exchanges, we multiply Eq.~(\ref{eqmotion1}) by $\partial d/\partial t$:
\begin{equation} \label{eqpower1}
m \frac{\partial^{2} d}{\partial t^{2}}  \frac{\partial d}{\partial t} +
\frac{\partial^{2}}{\partial x^{2}}
\left[ \left( k^{r}+\gamma \frac{\partial}{\partial t} \right)
\frac{\partial^{2} d}{\partial x^{2}} \right] \frac{\partial d}{\partial t}
= F \frac{\partial d}{\partial t}
\end{equation}
where we have dropped the explicit dependence over $x$ and $t$ to lighten notations.
The first term in the lhs of eq.~(\ref{eqpower1}) is the derivative of the local kinetic energy, while the term in the rhs is the power generated by the external force. To be able to identify the local dissipation of energy, one needs to rewrite eq.~(\ref{eqpower1}) under the general form of a balance equation, with a local time derivation, the divergence of an energy flux describing the energy transfer along the cantilever, and local dissipation and injection rates.
Hence, the second term in the lhs of eq.~(\ref{eqpower1}) needs to be transformed in order to be given a correct interpretation. Performing two integrations by parts, we get
\begin{equation} \label{eq:balance:td}
\frac{\partial }{\partial t} (\varepsilon_{\rm kin}+ \varepsilon_{\rm el})
+ \frac{\partial}{\partial x}\left( \Phi_{\rm el} + \Phi_{\rm vel} \right)
= W_{\rm diss} + W_{\rm ext}
\end{equation}
where $\varepsilon_{\rm kin}$ is the local kinetic energy,
$\varepsilon_{\rm el}$ is the local elastic energy,
$\Phi_{\rm el}$ and $\Phi_{\rm vel}$ are the energy fluxes through the cantilever
respectively due to elastic and viscoelastic forces,
$W_{\rm diss}(x,t)$ is the local dissipated power due to viscoelastic forces,
and $W_{\rm ext}(x,t)$ is the power injected by the external force.
These quantities are defined as
\begin{align}
\varepsilon_{\rm kin} &= \frac{1}{2} m \left( \frac{\partial d}{\partial t}\right)^2\\
\varepsilon_{\rm el} &= \frac{1}{2} k^{r}(x) \left( \frac{\partial^2 d}{\partial x^2} \right)^2\\
\Phi_{\rm el} &= \frac{\partial d}{\partial t} \frac{\partial}{\partial x}
\left[ k^{r}(x) \frac{\partial^2 d}{\partial x^2} \right]
- k^{r}(x) \left( \frac{\partial}{\partial t} \frac{\partial d}{\partial x} \right)
\frac{\partial^2 d}{\partial x^2}\\
\Phi_{\rm vel} &= \frac{\partial d}{\partial t} \frac{\partial}{\partial x}
\left[ \gamma(x) \frac{\partial}{\partial t} \frac{\partial^2 d}{\partial x^2} \right]
- \gamma(x) \left( \frac{\partial}{\partial t} \frac{\partial d}{\partial x} \right)
\frac{\partial}{\partial t} \frac{\partial^2 d}{\partial x^2}\\
\label{eq:Wdiss}
W_{\rm diss} &= \gamma(x) \left( \frac{\partial}{\partial t}  \frac{\partial^2 d}{\partial x^2} \right)^2\\
W_{\rm ext} &= F(x,t) \frac{\partial d}{\partial t}
\end{align}
Thanks to this decomposition, we have thus been able to determine the local (time-dependent) dissipation rate, which is given by eq.~(\ref{eq:Wdiss}). Replacing in eq.~(\ref{eq:Wdiss}) the deflection $d$ by the normal mode $\phi_n$ and averaging over time, one recovers after proper normalization the expression (\ref{eq:wdiss:normAppendix}) of the normalized dissipation rate (we recall that frequency dependences are neglected when we focus on a single mode).

\end{document}